\documentclass{article}

\usepackage[T1]{fontenc} 
\usepackage[utf8]{inputenc}
\usepackage{csquotes}
\usepackage{microtype}
\usepackage{authblk}
\usepackage{endnotes}
\usepackage{import}
\usepackage{mathtools}
\usepackage{hyperref} 
\usepackage{url}
\usepackage{natbib}

\hypersetup{
	pdfauthor={Paul Waddell, Ignacio Garcia-Dorado, Samuel M. Maurer, Geoff Boeing, Max Gardner, Emily Porter, Daniel Aliaga},
	pdftitle={Architecture for Modular Microsimulation of Real Estate Markets and Transportation},
	pdfsubject={Architecture for Modular Microsimulation of Real Estate Markets and Transportation},
	pdfkeywords={street networks,microsimulation,urban modeling,transportation,urban planning,regional planning,computer science,land use},
	pdffitwindow=true,         
	breaklinks=true,           
	colorlinks=false,          
	pdfborder={0 0 0}          
}

\let\footnote=\endnote

\title{Architecture for Modular Microsimulation of Real Estate Markets and Transportation\footnote{Paper presented at the Symposium on Applied Urban Modelling. Cambridge, England. June 27--29, 2018.}}

\author[1]{Paul Waddell}
\author[2]{Ignacio Garcia-Dorado}
\author[1]{Samuel M. Maurer}
\author[1]{Geoff Boeing}
\author[3]{Max Gardner}
\author[3]{Emily Porter}
\author[4]{Daniel Aliaga}
\affil[1]{Department of City and Regional Planning, University of California, Berkeley}
\affil[2]{Google Research}
\affil[3]{Department of Civil and Environmental Engineering, University of California, Berkeley}
\affil[4]{Department of Computer Science, Purdue University}

\date{June 2018}

\begin{document}

\maketitle

\begin{abstract}
Integrating land use, travel demand, and traffic models represents a gold standard for regional planning, but is rarely achieved in a meaningful way, especially at the scale of disaggregate data. In this paper, we present a new architecture for modular microsimulation of urban land use, travel demand, and traffic assignment. UrbanSim is an open-source microsimulation platform used by metropolitan planning organizations worldwide for modeling the growth and development of cities over long ($\sim$30 year) time horizons.  ActivitySim is an agent-based modeling platform that produces synthetic origin--destination travel demand data, developed from the UrbanSim model and software framework. For traffic assignment, we have integrated two approaches. The first is a static user equilibrium approach that is used as a benchmark.  The second is a traffic microsimulation approach that we have highly parallelized to run on a GPU in order to enable full-model microsimulation of agents through the entire modeling workflow. This paper introduces this research agenda, describes this project's achievements so far in developing this modular platform, and outlines further research.
\end{abstract}

\newpage

\section{Introduction}
\label{sec:intro}
\subsection{Need for modular urban microsimulation}

Integrated urban models have a long history and uneven results.  \cite{lee-1973} criticized early urban models as committing seven deadly sins, among them being too data hungry.  In the intervening years much has changed, including our conceptions of big data, and the planning challenges models are being tasked to address.  Reduction of greenhouse gas emissions and vehicle miles travelled are critical objectives for metropolitan planning in California and other states, but these objectives increasingly need to be balanced with housing affordability and environmental justice objectives, among many others.  The policy and planning landscape has also become much more diverse and sophisticated, with complex variations of traditional zoning, form-based codes, inclusionary zoning, parking requirements, density bonuses, impact fees and subsidies, and community benefits, among many others, now being used to influence development patterns at multiple scales, with considerable emphasis on the site and walking scale.

In this project we attempt to address these evolving needs by developing a  modular  microsimulation architecture based on UrbanSim (Waddell, 2002, 2011) that uses parcels and local street and transit networks, along with representations of individual persons, households, jobs and business establishments to model the dynamic evolution of the San Francisco Bay Area and its daily travel patterns.  We associate households with individual residential units that have characteristics such as tenure, square footage, bedrooms, bathrooms. The units are associated with buildings, which in turn are located on parcels, and are linked to the local street network.  We analyze accessibility on local street networks, transit networks and driving networks, and microsimulate traffic patterns and congestion using a parallelized traffic microsimulation model that uses the many cores on a graphics card to parallelize its processing~\citep{garcia2014designing}. Where possible, the integrated model takes advantage of a new templating system for generating model steps that fit directly into an orchestration pipeline \citep{maurer-2018}. 

Our goal is to enable high-resolution modeling at a high enough level of performance to support analysis of local and complex land use policies, in conjunction with transportation modeling that enables full treatment of walking, biking and transit on the same basis as driving, for the first time.  Prior models have privileged driving over other modes by suppressing the spatial detail needed to adequately represent the walking, biking and even transit options realistically. 

The project is funded by the Department of Energy and is part of a larger Smart Mobility project that is intended to anticipate the impacts of Transportation Network Companies (TNCs), such as Lyft and Uber, and the rapidly evolving technology for autonomous vehicles (AVs), the potential impacts of which are relatively poorly understood.  Our modeling platform is intended to provide a flexible framework to support extension as data emerge on the adoption of TNC and eventually AV technologies, and to examine their consequences on urban development and travel patterns, on energy consumption, and on unintended social and environmental consequences.

In this paper we describe the architecture of the platform, the tools and data developed to implement it, and initial results from the model implementation.  We also address the limitations of the model, and outline further steps needed to refine it to a point it can be brought into broader practical use.  The paper focuses on the model system architecture rather than on specific empirical results, and is being developed and empirically tested in the San Francisco Bay Area initially, with the intent to scale it to  other metropolitan areas in the United States and internationally.

The overarching objective of this project is to develop a modular modeling pipeline that encompasses land use, travel demand, and traffic assignment to model the combined and cumulative impacts of transportation infrastructure and land use regulations. A key motivation for developing such a model system is that the urban environment
is complex enough that it is not feasible to anticipate the effects of alternative infrastructure investments and land use policies without some form of analysis that could reflect the cause and effect interactions that could have both intended and unintended consequences.  The emphasis on modularity is motivated by a desire to implement an architecture that lends itself to rapid evolution and innovation, enabling model components to follow some straightforward design patterns in order to quickly substitute alternative models and evaluate them easily.

In order to better support the analysis of the impacts of transportation infrastructure and land use regulations within large and complex urban regions, we propose to develop an integrated pipeline for modeling urban land use, travel demand and traffic assignment and to compute transportation-related energy consumption. The project lends itself to further extension to address building energy consumption as well, creating the potential to coherently simulate transport and building energy demand for the first time in a coherent way, at an urban and metropolitan scale.

\subsection{Overview of modular microsimulation architecture}

Three existing models are leveraged within this project. UrbanSim is a model system developed to represent long-term dynamics of urban development and its interaction with transportation systems. ActivitySim is an activity-based travel demand model system, and it was developed using the UrbanSim platform as its starting point. The third model component is a static user equilibrium traffic assignment model using a standard Frank-Wolfe algorithm.  All three model systems are open source and implemented using the Python programming language, enabling broad collaboration within the research community and the public agencies who could benefit from its use.  Computational performance using Python is achieved in UrbanSim and ActivitySim using vectorized calculations with math libraries such as Numpy that are implemented in C, thus avoiding the performance penalties of iterative processing in Python. In addition to the static user equilibrium traffic assignment model, we also leverage a traffic microsimulation model implemented on a graphics processing unit (GPU).

\begin{figure}[htbp]
  \center
  \includegraphics[width=\textwidth]
  {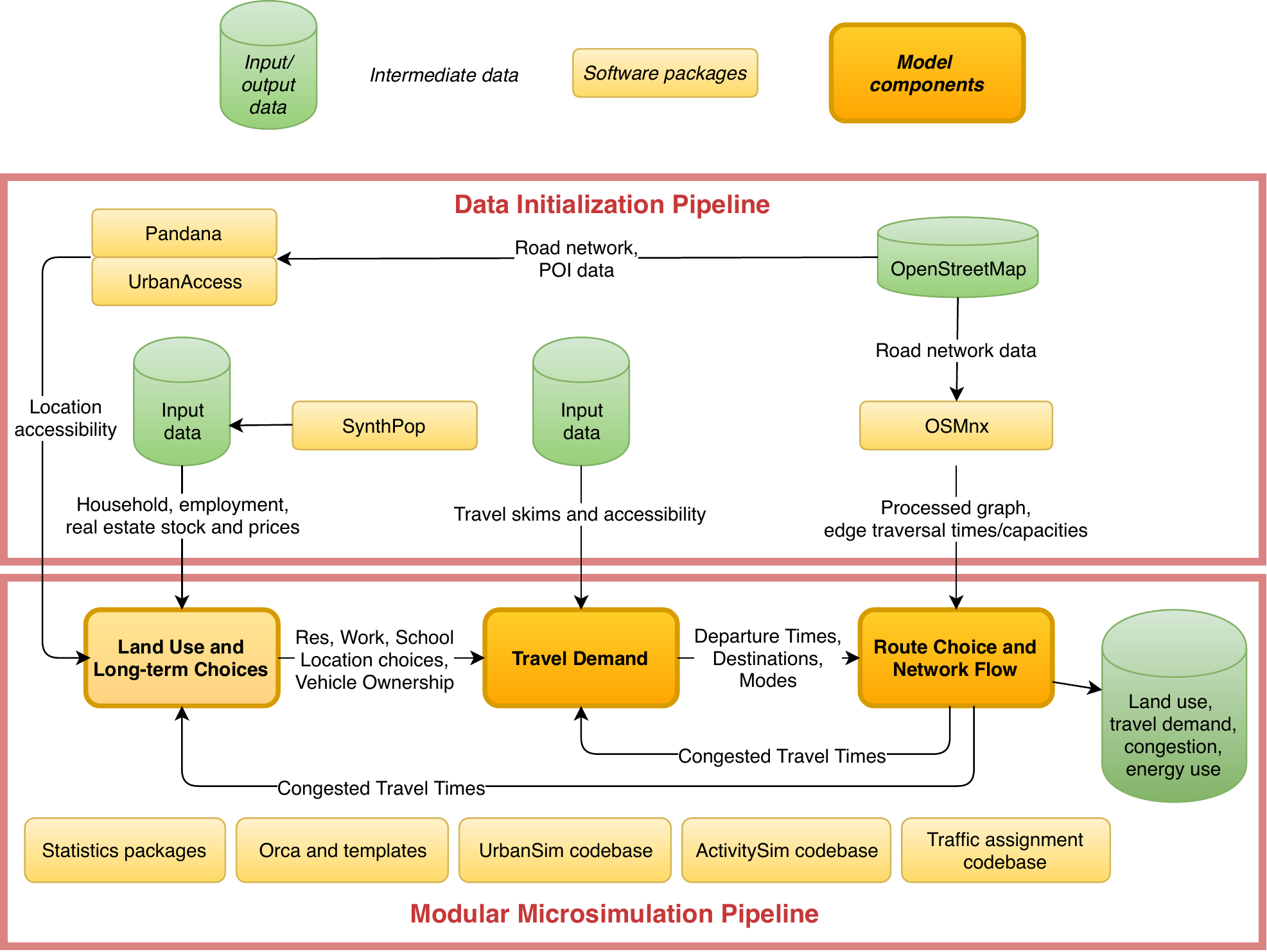}
  \caption{Overview of the integrated modeling pipeline's modular architecture}
  \label{fig:overview_pipeline_architecture}
\end{figure}

While UrbanSim and ActivitySim are microsimulation models, meaning that they operate at the level of individuals and households, the traffic assignment model is an aggregate traffic flow model. Figure \ref{fig:overview_pipeline_architecture} depicts the pipeline for integrating these three models.

The integrated model presented in this paper is the first to make use of a new template system for generating model steps that fit directly into an orchestration pipeline. Common types of model steps that use ordinary least squares regression, binary logit, or various forms of multinomial logit can be generated directly from an API, enabling easier model development, iteration, and validation. This template system is described in more depth in \cite{maurer-2018}.

\section{Long-term land use model: UrbanSim}
\label{sec:urbansim}
\subsection{Overview}

UrbanSim has been developed to support land use, transportation and environmental planning, with particular attention to the regional transportation planning process \citep{waddell-japa-2002, waddell-tra-2007, waddell-tr-2011}. It has also recently adapted to address housing affordability concerns and the potential for displacement around transit investments and upzoning associated with transit oriented development, and is being extended to support the evaluation of planning for adaptation to or mitigation of hazards associated with climate change or earthquakes. 

\subsection{Inputs}

UrbanSim uses a detailed representation of the built environment and its occupants.  It implements a microsimulation data structure, meaning that it represents every household, person, job, parcel, and building in a metropolitan area. The population data is generated from census data using synthetic population algorithms we have developed \citep{ye-trb-2009}.  The employment data is from an inventory of business establishments compiled by the Metropolitan Planning Organization from commercial data sources such as InfoGroup, or from state unemployment insurance records. Parcel and building data are also generally obtained by the MPO, generally from the individual counties and cities in the region.

In addition, land use regulations from  municipalities within a metropolitan area are assembled and reconciled into a regional land use regulation database, and a database of development projects \enquote{in the pipeline} for development was also assembled, to be able to accurately reflect development that is in progress.  Real estate prices and rents are obtained from county assessor records, or from commercial data source such as CoStar for use in estimating the rent and price models in UrbanSim.  Finally, household travel survey data is used in estimating the Household Location Choice Model in UrbanSim.

Beyond these base data sets, the other main input to UrbanSim is what we refer to as scenario inputs. We use the term \emph{scenario} in the context of UrbanSim in a very specific way: a scenario is a combination of input data and assumptions to the model system, including macroeconomic assumptions regarding the growth of population and employment in the study area, the configuration of the transportation system assumed to be in place in specific future years, and general plans of local jurisdictions that will regulate the types of development allowed at each location.

In order to facilitate comparative analysis, a model user such as a Metropolitan Planning Organization will generally adopt a specific scenario as a base of comparison for all other scenarios. This base scenario is generally referred to as the \emph{baseline} scenario, and this is usually based on the adopted or most likely to be adopted regional transportation plan, accompanied by the most likely assumptions regarding economic growth and land use policies. Once a scenario is created, it determines several inputs to UrbanSim:

\begin{itemize}
    \item \textit{Control totals}: data on the aggregate amount of population and employment, by type, to be assumed for the region.
    \item \textit{Travel data}: data on zone to zone travel characteristics, from the travel model.
    \item \textit{Development constraints}: a set of rules that interpret the general plan codes, to indicate the allowed land use types and density ranges on each parcel.
\end{itemize}

\subsection{How it works}

UrbanSim predicts the evolution of household residential locations, firm locations, buildings, prices and rents over time, using annual steps. It is interfaced with a metropolitan travel model system to deal with the interactions of land use and transportation. 
Accessibility is a well explored area of urban theory, and is the concept that connects transportation and land use, so is is central to understanding the UrbanSim model system.  

Operationally used accessibility frameworks include gravity-model based (defined by attractions and discounted by distance), cumulative-opportunity (summations within a set impedance measure) and space-time (limited by the opportunity prism of an individual's activity skeleton) \citep{kwan_space-time_1998,miller_measuring_1999}.  Dong and others expand on the space-time prisms by creating a logsum-based measure within a travel model \citep{dong_moving_2006}.  

To measure access, one must first choose a basic unit of geography to use.  The majority of transportation models in use today still rely heavily on zone-based geography for its simplicity and computational tractability. 
Zones can vary in size, but are usually a few city blocks at their smallest.  Drawbacks to this method include: the zones must be defined manually, they are arbitrary in scope, and they are too large to model micro-land use measures and walkability (which often vary on a block-by-block basis). We have developed algorithms to support walking scale queries on local street networks that enable fast accessibility queries to be computed on metropolitan scale networks \citep{foti2012generalized}.  In subsequent work we have generalized these algorithms to address walking plus transit networks \citep{blanchard-waddell-trr-2017, blanchard2017urbanaccess}. 



UrbanSim makes extensive use of models of individual choice. A  model of households choosing among alternative locations in the housing market, provides an illustrative example. For each agent, we assume that each alternative location $i$ has associated with it a utility $U_i$ that can be separated into a systematic part and a random part:

\begin{equation}
    \label{eq:utility}
    U_i = V_i + \epsilon_i
\end{equation}

where $V_i = \beta\cdot {x}_i$ is a linear-in-parameters function, $\beta$ is a vector of $k$ estimable coefficients, $x_i$ is a vector of observed, exogenous, independent alternative-specific variables that may be interacted with the characteristics of the agent making the choice, and $\epsilon_i$ is an unobserved random term. Assuming the unobserved term in Equation \ref{eq:utility} to be distributed with a Gumbel distribution leads to the widely used multinomial logit model \citep{mcfadden-1974,mcfadden-1981}:

\begin{equation}
    \label{eq:mnl}
    P_i = \frac{\mathrm{e}^{V_i}}{\sum_j \mathrm{e}^{V_j}}
\end{equation}

where $j$ is an index over all possible alternatives. The estimable coefficients of Equation \ref{eq:mnl}, $\beta$, are estimated using maximum likelihood \citep{greene-2002}.

In order to balance demand and supply in the short run, we developed a choice algorithm that enables the model to simulate short-term market clearing processes. We compute the probability step of the location choice model, sum the probabilities at each submarket to compute aggregate demand, and use this estimate of demand to compare to the available supply in the submarket. Prices are adjusted iteratively and the relevant components of the location choice model are updated to reflect the influence of the adjusted prices. This algorithm captures the feedback loop between excess demand for locations causing prices there to increase, which in turn, dampens demand as the submarket becomes relatively more expensive than other submarkets that are substitutes. 


\subsection{Outputs}

As a microsimulation system, UrbanSim  produces the same outputs as it uses as inputs for a future point in time: tables of individual households and persons, jobs, parcels, buildings, with their attributes updated each simulation year if they have been modified by the model system.  New households, jobs and buildings are added or subtracted by the simulation, and households and jobs may relocate into or within the region.

By retaining this level of detail, UrbanSim is able to generate summaries of the real estate data, demographics, or economic profile of any geographic aggregation requested by the user, such as census geographies, cities, counties, or other planning geographies.  Often traffic analysis zone summaries are used as inputs to the travel model system.  In this project our goal is to avoid losing information by aggregating the data to traffic zone when we connect UrbanSim to the travel demand model system.

\subsection{Calibration and validation}

UrbanSim is generally calibrated longitudinally, starting from an observed year in the past and running it to a later observed year, to compare predicted to observed data over time, and adjusting calibration coefficients iteratively to improve the fit of the model to the observed calibration targets at the calibration year.  Generally it is preferable to minimize the use of calibration constants since excessive use of calibration constants can \enquote{handcuff} the model and make it insensitive to policy changes.

Some work has been done previously to extend this methodology to account for uncertainty using Bayesian Melding, to calibrate the model uncertainty and enable the computation of confidence intervals around its predictions, when running the model multiple times without fixing the random seed for the stochastic simulation \citep{sevcikova-tra-2009, sevcikova-tra-2011}.

\subsection{Benchmarking}

In our tests, a single UrbanSim iteration for a synthetic population of $\sim$2.6M households and $\sim$6.9M individuals completed in just under 11 minutes of wall-clock time on a Ubuntu Linux machine with 24 Intel Xeon X5690 3.47GHz CPUs. Extrapolating to the typical 30-year scenario, a complete UrbanSim run will take $\sim$5.5 hours using one-year intervals or $\sim$1.1 hours using a five-year interval.

\section{Short-term travel demand model: ActivitySim}
\label{sec:activitysim}
\subsection{Overview}

ActivitySim is an agent-based modeling (ABM) platform for modeling travel demand. Like UrbanSim, the ActivitySim software is entirely open source, and hosted as a part of the Urban Data Science Toolkit\footnote{The open-source Urban Data Science Toolkit is available online at \url{https://github.com/UDST}}. ActivitySim grew in large part out of a need for metropolitan planning organizations (MPOs) to standardize the modeling tools and methods that were common between them in order to facilitate more effective collaboration and sharing of innovations.

Today, ActivitySim is both used and maintained by an active consortium of MPOs, transportation engineers, and other industry practitioners. Because of the cooperative approach taken by ActivitySim stakeholders towards its ownership, and because many of its \enquote{owners} are also its main users, the platform continues to mature in the direction that most benefits the practitioners themselves. ActivitySim development is still in beta, with an official 1.0 release scheduled for 2018.

\subsection{Inputs}

ActivitySim requires two main sets of input data, one relating to geography and the other relating to the population of synthetic agents whose travel choices are being modeled.

The geographic data are stored at the level of the traffic analysis zone (TAZ) and are comprised of three components: 1) land use characteristics; 2) a matrix of zone-to-zone travel impedances (travel times, distances, or costs) specific to the mode of travel and time of day; and 3) a table of user-defined measures of aggregate utility estimated for each zone. In transportation planning, these zone-level impedances and utility measures are commonly referred to as \emph{skims} and \emph{accessibilities}, respectively.

The land use data consist of zone-level population and employment characteristics, along with measures of different land use and building types. In our integrated model these data are read directly from the outputs generated by UrbanSim, but for a single simulation iteration any source of aggregate land use data would suffice.

Travel skims are typically generated by a traffic assignment model, which ActivitySim is not. ActivitySim instead expects to load the skims from an OpenMatrix (OMX) formatted data file\footnote{The OpenMatrix format is specified online at \url{https://github.com/osPlanning/omx/wiki}}. The creation of these skims is described below in Section \ref{sec:ta} on traffic assignment.

Accessibilities can be generated directly from the skims or any other graph representation of the transportation network. They are computed by aggregating mode-specific measures of access to specific amenity types across the network, most commonly employment centers, retail outlets, and transportation hubs. The measures of access can be as simple as counts of amenities reachable within a given shortest-path distance, or as complex as composite utilities generated by a discrete choice model. 

The second set of ActivitySim input data is the synthetic population. The synthetic population data consist of both individuals and their characteristics, as well as the households and household characteristics into which the individuals are organized. The synthetic population is shared between UrbanSim and ActivitySim, although UrbanSim does not make use of individual-level characteristics.

The exhaustive details of the ActivitySim data schema are documented online\footnote{The ActivitySim data schema is available online at \url{https://udst.github.io/activitysim/dataschema.html}}.

\subsection{How it works}

ActivitySim, like UrbanSim, relies heavily on discrete choice models and random utility maximization theory \citep{mcfadden-1974}. Please refer to Section \ref{sec:urbansim} for specific details about how discrete choice models work within an agent-based microsimulation framework.

\begin{figure}[htbp]
    \center
    \includegraphics[width=\textwidth]{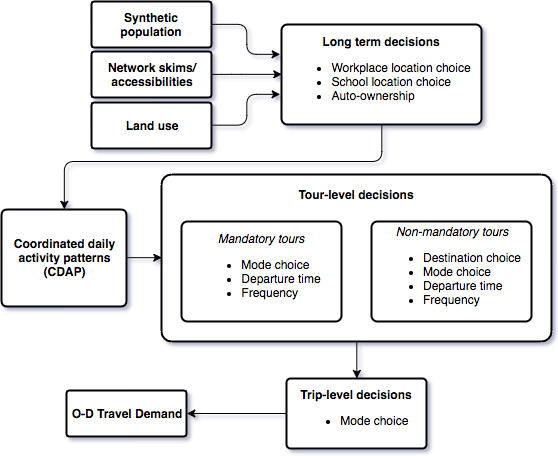}
    \caption[ActivitySim model flow]{ActivitySim model flow\footnote{The ActivitySim model flow is adapted from \url{http://analytics.mtc.ca.gov/foswiki/bin/view/Main/ModelSchematic}}}
    \label{fig:asim-models}
\end{figure}

An ActivitySim run consists of a series of sequentially executed model steps. The individual models can be grouped into the four clusters---long term decisions, coordinated daily activity patterns, tour-level decisions, and trip-level decisions---illustrated in Figure \ref{fig:asim-models} and summarized briefly here:

\begin{itemize}
    \item \emph{Long-term choice models}: ActivitySim's three long-term choice models---workplace location choice, school location choice, and auto-ownership---model the choices that are not made every day in the real world but have a substantial impact on those that are. These models will eventually be migrated to run directly in the UrbanSim environment so that the time horizons of the two simulation platforms are internally consistent.
    
    \item \emph{Coordinated Daily Activity Patterns}: the CDAP step models the group decision-making process for individual household members all seeking to maximize the utility of their daily activities together. CDAP takes into consideration mandatory and non-mandatory trips choosing activities to maximize each individual's utilities. The maximization process currently involves the estimation of all possible combinations of all individuals within a household, and thus has the longest run-time of all ActivitySim models.
    
    \item \emph{Tour-level decisions}: tours define chains of trips that are completed together without returning home in between. Mandatory tours include trips to and from work and school, while non-mandatory trips are entirely discretionary. Non-mandatory tour alternatives are specified in a user-defined configuration file, and thus these steps include a destination choice model as well. Mandatory tour alternatives have already been computed by the long-term decision models. Each tour type has separate model steps for estimating mode choice, departure time, and the frequency of the tour.
    
    \item \emph{Trip-level decisions}: mode choice must be selected at the level of the individual trip as well as the tour because a given tour may include different modes for different trip legs. Trip departure and arrival times are estimated as well. The rest of the trip characteristics are inherited from the tours to which a trip belongs. 
\end{itemize}

\subsection{Outputs}

The output of an ActivitySim run consists of a single HDF5 data file with a single table of results corresponding to each model step, along with the versions of the input files in their final, updated states. For the purpose of generating travel demand for traffic assignment, however, we are only concerned with the output of the trip generation step. This single file contains the origin and destination zones, start and end times, and mode choice for every trip taken by every agent over the course of a day. We then take the subset of these trips that are completed by automobile and aggregate the counts by origin-destination pair and hour of departure. These hourly, zone-level demand files are finally handed off for use in traffic assignment.

\subsection{Calibration and validation}
Compared to their meso- and macro-scale counterparts, microsimulations like ActivitySim more accurately capture the nonlinearities that define most patterns of human behavior by modeling the decision-making processes of individual agents. The models themselves, however, are not meant to be interpreted on the same disaggregate scale. We do not know which individuals will use which mode to complete which activity on a given day, but rather how an entire population of individuals is likely to behave \textit{en masse}.

As such, there are a variety of data sets available to us for validating our results, including the Bay Area Travel Survey (BATS), the U.S. Census Longitudinal Employer-Household Dynamics program (LEHD), and the California Household Travel Survey (CHTS). All of these products offer data that can be aggregated to the TAZ level and compared to the output of our models.

\subsection{Benchmarking}

In our tests, a full ActivitySim run for a synthetic population of $\sim$2.6M households and $\sim$6.9M individuals completed in just over 16 hours of wall-clock time on a Ubuntu Linux machine with 24 Intel Xeon X5690 3.47GHz CPUs. Over one-third of the runtime is accounted for by the CDAP step, and future work should concentrate on either improving the runtimes of this step, or implenting a bypass step in which intra-household decisions are vastly simplified.

\section{Road network model}
\label{sec:network}
\subsection{Overview}

Once we have produced this synthetic travel demand data, we model the regional circulation network for subsequent traffic assignment or simulation. Our integrated pipeline models networks as mathematical graphs consisting of a set of nodes $N$ connected to one another by a set of edges $E$ \citep{newman_networks:_2010,gastner_spatial_2006}. Specifically, the road network is modeled as a nonplanar directed multigraph with possible self-loops. The data come from OpenStreetMap.

This section describes how we acquire these data, construct a graph model of the network, process its topology, and infer and impute relevant variables. Next it describes the process of calculating BPR coefficients for static assignment---and the assumptions baked into these calculations. Finally, it explains the process of converting the zone-based travel demand output from ActivitySim to a network node-based demand data set for zone-less modeling.

\subsection{Network creation, construction, and processing}

The network data in this modeling pipeline come from OpenStreetMap. OpenStreetMap is a public worldwide collaborative mapping project and web platform with over two million users. Anyone may edit or access OpenStreetMap data, but community oversight and standards exist to prevent significant vandalism or inaccurate edits \citep{jokar_arsanjani_openstreetmap_2015}. In general, OpenStreetMap data are of high accuracy and quality, particularly in the United States and Western Europe \citep{corcoran_analysing_2013,over_generating_2010,haklay_how_2010,maron_how_2015}. OpenStreetMap imported the 2005 TIGER/Line roads data set as a foundation, and numerous corrections and additions to these data have been made since \citep{willis_openstreetmap_2008}.

\begin{figure}[htbp]
    \center
    \includegraphics[width=\textwidth]
    {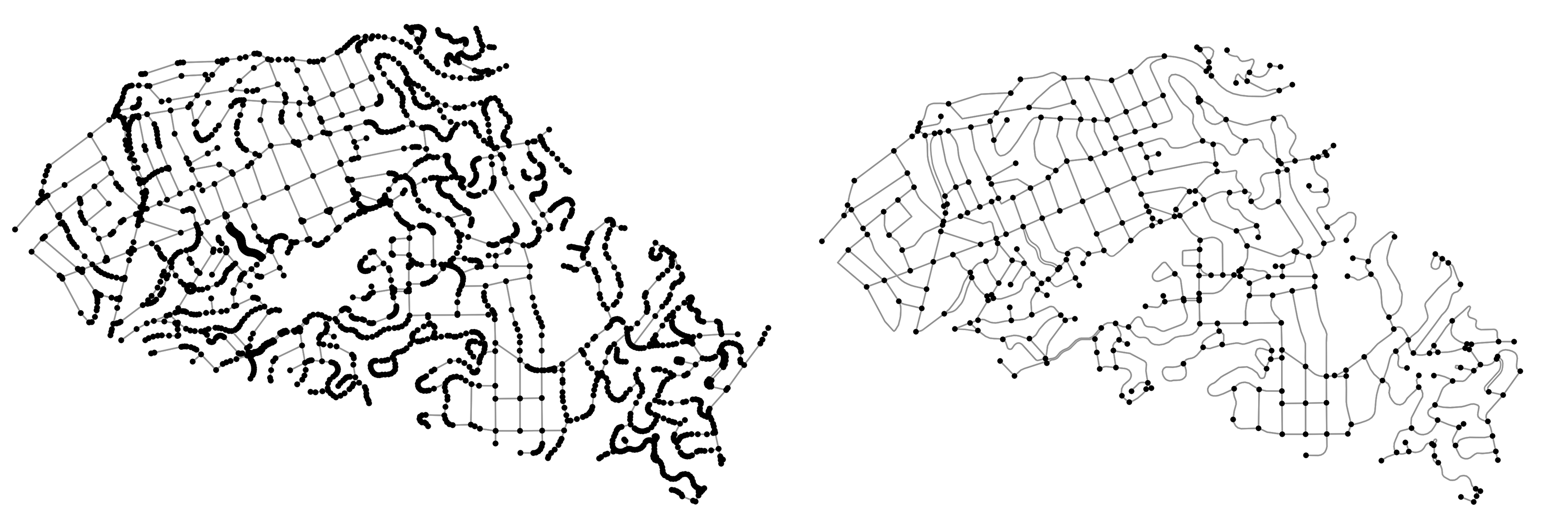}
    \caption{A suburban street network from OpenStreetMap data before OSMnx topology simplification (left) and after (right), with nodes in black and edges in gray. Note that the full spatial geometry is retained as edge metadata even though the edge itself is compressed to a single pair of origin and destination nodes.}
    \label{fig:simplification_before_after}
\end{figure}

The road network graph is constructed using OSMnx, an open-source Python package for working with OpenStreetMap data \citep{boeing_osmnx:_2017}. OSMnx is built on top of NetworkX, a Python package for network analysis developed by researchers at Los Alamos National Laboratory. OSMnx extends NetworkX's network analysis capabilities by working explicitly with spatial infrastructure networks and interfacing with OpenStreetMap's various APIs. It can automatically construct topologically-simplified nonplanar directed multigraphs constrained to any polygonal boundaries for anywhere in the world from OpenStreetMap data \citep{boeing_multiscale_2018}. OSMnx uses a multistep algorithm to simplify the topology of the graph so that it retains nodes only at intersections and dead-ends, as well as the full geometry of the simplified edges, as shown in Figure \ref{fig:simplification_before_after}.

We use OSMnx to download the road network for the nine-county San Francisco Bay Area. We use the 2016 US Census Bureau TIGER/Line shapefile of United States counties to define the spatial extents of these nine counties: Alameda, Contra Costa, Marin, Napa, San Francisco, San Mateo, Santa Clara, Solano, and Sonoma. We calculate a convex hull around these geometries to obtain a single polygon for the spatial query---this prevents the query from discarding any network elements that fall within our study area but outside of the counties' official borders (for example, bridges over the San Francisco Bay).

Next, we use OSMnx to download the drivable road network within this convex hull spatial boundary. OSMnx processes the detailed metadata tags to identify which paths are drivable. It then constructs them into an in-memory graph (a multidigraph to be precise). This initial graph contains 1.2 million nodes and 2.3 million edges.

Next we filter the road network to retain only tertiary roads and higher. In OpenStreetMap terminology, the road types we retain comprise: motorway, motorway\textunderscore link, trunk, trunk\textunderscore link, primary, primary\textunderscore link, secondary, secondary\textunderscore link, tertiary, tertiary\textunderscore link, unclassified, and road. The \enquote{link} types are necessary to retain the connectors (such as on-ramps and off-ramps) between certain roads. The \enquote{road} type is standardly used in the OpenStreetMap community as a null value. The \enquote{unclassified} type technically refers to the British-style roads hierarchy, in which \enquote{unclassified} is the level below \enquote{tertiary}---that is, essentially, quaternary. However, it is sometimes used inconsistently in the United States, so we retain it for completeness.

\begin{figure}[htbp]
    \center
    \includegraphics[width=\textwidth]
    {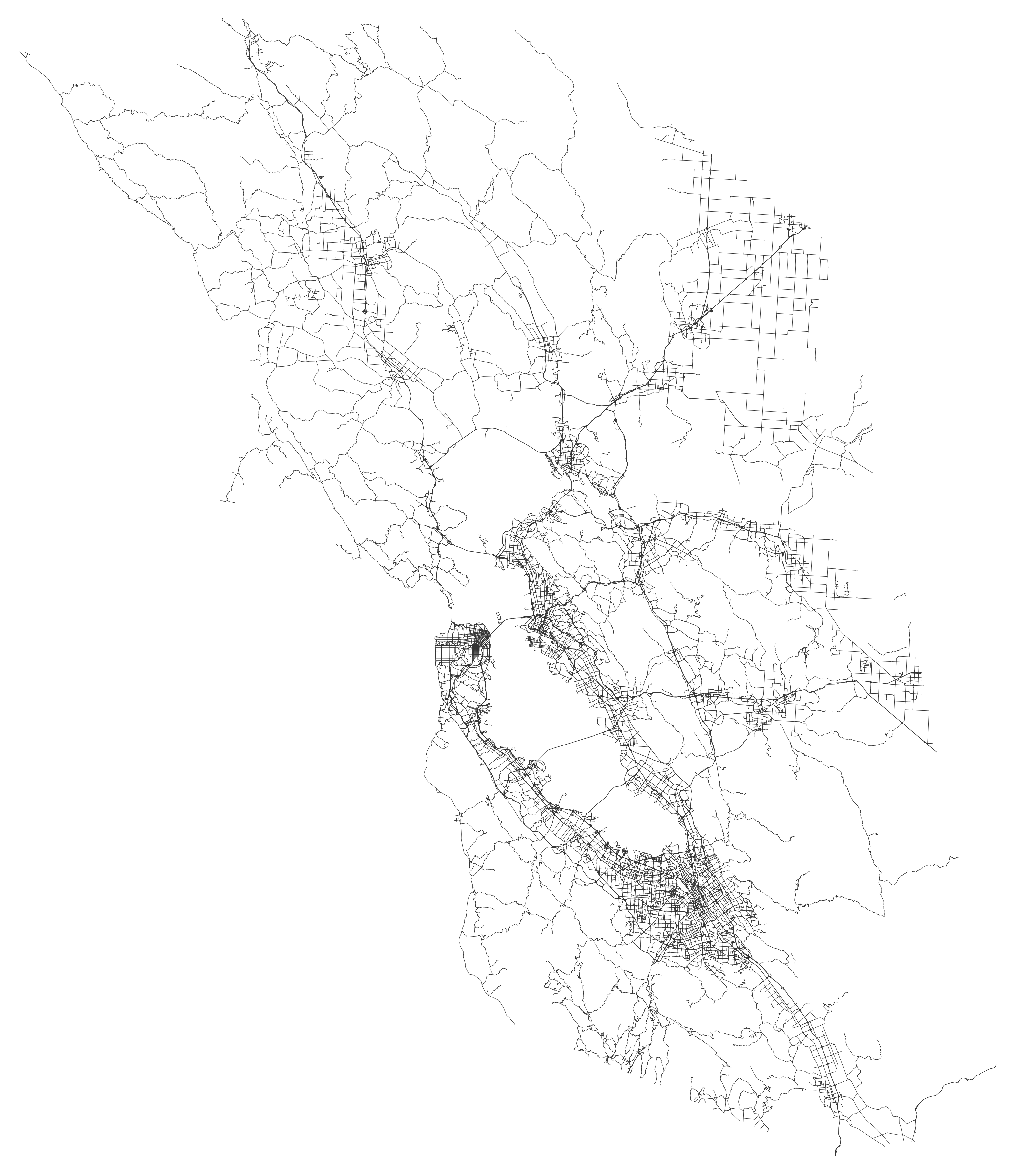}
    \caption{The nine-county San Francisco Bay Area road network used in our modeling pipeline, created by OSMnx from OpenStreetMap data.}
    \label{fig:bay_area_road_network}
\end{figure}

After filtering out all edges not of these major types, we remove all isolated nodes then retain only the largest strongly-connected component of the graph. Finally, we simplify the graph using OSMnx then save it to disk as a GraphML file as well as a nodes list and an edges list. GraphML is a standard, XML-based file format for storing and exchanging complete graph structure data. Our final graph contains 31,000 nodes and 66,000 directed edges representing 20,000 kilometers of roads across these 9 San Francisco Bay Area counties, as seen in Figure \ref{fig:bay_area_road_network}.

\subsection{BPR coefficients calculation and assumptions}

As this modeling pipeline is modular and can support various traffic assignment/simulation methodologies, we process the graph node and edge attribute data to make them suitable for modeling traffic via static user equilibrium assignment, dynamic assignment, and vehicle microsimulation approaches.

To define the relationship between edge travel time and edge congestion we use the Bureau of Public Roads (BPR) congestion function. This BPR function has long been used by transportation engineers to model the increased time needed to traverse an edge when congestion on the edge increases. Once we have the final processed graph of the Bay Area road network, we calculate the BPR  curves' coefficients for each edge. The BPR function is defined in Equation \ref{eq:bpr_function}.

\begin{equation}
    t_i = t^0_i (1 + \alpha (\frac{v_i}{c_i}) ^ \beta)
    \label{eq:bpr_function}
\end{equation}

where:

\begin{itemize}
    \item $t_i$ is the congested flow travel time on edge $i$
    \item $t^0_i$ is the free-flow travel time on edge $i$
    \item $v_i$ is the number of vehicles on edge $i$ per unit of time 
    \item $c_i$ is the capacity (i.e., maximum number of vehicles) of edge $i$ per unit of time
    \item $\alpha$ linearly increases the congested travel time with regards to the volume:capacity ratio
    \item $\beta$ exponentially increases the congested travel time with regards to the volume:capacity ratio
\end{itemize}

The $\alpha$ parameter was assigned a value of 0.15 in the original BPR curve, and the $\beta$ parameter was assigned a value of 4 in the original BPR curve. We adopt these default $\alpha$ and $\beta$ parameter values in this modeling pipeline, but note that they are not the only way to parameterize this function. For instance, the TRB's NHCRP Report 716: Travel Demand Forecasting Parameters and Techniques provides various coefficients estimated using the 1985 Highway Capacity Manual \citep[p.~75]{transportation_research_board_highway_1985,transportation_research_board_travel_2012}.

To calculate the BPR coefficients, we require per-edge data about capacity and free-flow travel time. To assemble these data, we require information about edge lengths, free-flow speed, and number of lanes. We use OSMnx to calculate these edge lengths in meters. OpenStreetMap contains sparse data per-edge on maximum permitted speed and number of lanes. When these data are missing, we infer or impute them as per the defaults\footnote{The authors wish to thank Madeleine Sheehan and Alexander Skabardonis for providing some of these values.} presented in Tables \ref{table:free_flow_speed_defaults} and \ref{table:capacity_defaults}.

First, for each edge that is missing number of lanes data, we impute the value based on its edge type (which OpenStreetMap always provides as metadata). We calculate this imputed value by taking the median value of all edges of this type. Second, for each edge that is missing its maximum permitted speed data, we infer the free-flow speed from a look-up table via edge type and number of lanes (see Table \ref{table:free_flow_speed_defaults}). We can now calculate free-flow travel time per edge as a function of free-flow speed and edge length, as defined in Equation \ref{eq:free_flow_travel_time}:

\begin{equation}
    t^0_i = \frac{d_i}{s_i}
    \label{eq:free_flow_travel_time}
\end{equation}

where:

\begin{itemize}
    \item $t^0_i$ represents free-flow travel time on edge $i$, in units of seconds
    \item $d_i$ represents the length of edge $i$, in units of meters
    \item $s_i$ represents free-flow speed (i.e. the maximum permitted speed of travel) on edge $i$, in units of meters per second
\end{itemize}

Next, we infer each edge's vehicle capacity (per lane per hour) from a look-up table via edge type and number of lanes (see Table \ref{table:capacity_defaults}). We then convert this capacity per lane per hour value to units of capacity per edge per second. Finally, we use these data to calculate the values of the $a0$ and $a4$ coefficients for the BPR curve. These calculations require a caveat: due to missing data on OpenStreetMap, we are forced to infer or impute variables on many edges. Our assumptions about parameter values and capacity and speed limit defaults inevitably propagate through to our BPR coefficients.

\begin{table}[htbp]
\centering
\caption{Free-flow speed defaults by edge type and number of lanes, in units of miles per hour.}
\label{table:free_flow_speed_defaults}
\begin{tabular}{ l r r r r } 
    \hline
    Edge type & lanes=1 & lanes=2 & lanes=3 & lanes=4+ \\
    \hline
    motorway & 50 & 50 & 65 & 65 \\
    motorway\textunderscore link & 50 & 50 & 65 & 65 \\
    trunk & 45 & 45 & 45 & 45 \\
    trunk\textunderscore link & 45 & 45 & 45 & 45 \\
    primary & 30 & 30 & 30 & 30 \\
    primary\textunderscore link & 30 & 30 & 30 & 30 \\
    secondary & 25 & 25 & 25 & 25 \\
    secondary\textunderscore link & 25 & 25 & 25 & 25 \\
    tertiary & 20 & 20 & 20 & 20 \\
    tertiary\textunderscore link & 20 & 20 & 20 & 20 \\
    unclassified & 20 & 20 & 20 & 20 \\
    road & 30 & 30 & 30 & 30 \\
    \hline
\end{tabular}
\end{table}

\begin{table}[htbp]
\centering
\caption{Capacity defaults by edge type and number of lanes, in units of vehicles per lane per hour.}
\label{table:capacity_defaults}
\begin{tabular}{ l r r r r } 
    \hline
    Edge type & lanes=1 & lanes=2 & lanes=3 & lanes=4+ \\
    \hline
    motorway & 1900 & 2000 & 2000 & 2200 \\
    motorway\textunderscore link & 1900 & 2000 & 2000 & 2200 \\
    trunk & 1900 & 2000 & 2000 & 2000 \\
    trunk\textunderscore link & 1900 & 2000 & 2000 & 2000 \\
    primary & 1000 & 1000 & 1000 & 1000 \\
    primary\textunderscore link & 1000 & 1000 & 1000 & 1000 \\
    secondary & 900 & 900 & 900 & 900 \\
    secondary\textunderscore link & 900 & 900 & 900 & 900 \\
    tertiary & 900 & 900 & 900 & 900 \\
    tertiary\textunderscore link & 900 & 900 & 900 & 900 \\
    unclassified & 800 & 800 & 800 & 800 \\
    road & 900 & 900 & 900 & 900 \\
    \hline
\end{tabular}
\end{table}

\subsection{Linking zone-based travel demand to the network}

The ActivitySim component of our integrated modeling pipeline produces an output dataset of zone-to-zone travel demand data. To model this travel demand (and in turn traffic assignment) on our road network, we must convert the trip origins and destinations from zones to network nodes.

We use the shapefile of Traffic Analysis Zones (TAZs) from the Bay Area Metropolitan Transportation Commission (MTC) to acquire zone spatial extents. Then we calculate the centroid of each zone polygon. Finally, we identify the network node nearest to each zone's centroid by taking the minimum of a vectorized calculation of great-circle distances from the centroid to every node in the network, using the haversine formula defined in Equation \ref{eq:haversine_formula}:

\begin{equation}
    \delta_{gc} = r\cdot \arccos{(\sin{\Phi_1}\cdot\sin{\Phi_2} + \cos{\Phi_1}}\cdot\cos{\Phi_2}\cdot\cos{|\lambda_1 - \lambda_2|})
    \label{eq:haversine_formula}
\end{equation}

where:

\begin{itemize}
    \item $\delta_{gc}$ represents the great-circle distance between the two points, in meters
    \item $r$ represents the radius of the Earth, in meters
    \item $\Phi_1$ and $\Phi_2$ represent the geographical latitudes of the two points, in radians
    \item $\lambda_1$ and $\lambda_2$ represent the geographical longitudes of the two points, in radians
\end{itemize}

Now that we have the road network, its per-edge BPR coefficients, and node-based travel demand data, we are ready to model traffic itself.

\section{Static traffic assignment model}
\label{sec:ta}
\subsection{Overview}

The traffic assignment component of this integrated modeling pipeline provides vehicles in the network specific paths to make trips between their origins and destinations (ODs). The origins, destinations, and numbers of vehicles making trips are received from ActivitySim and then assigned to specific network edges through traffic assignment. The resulting path assignment results in a total number of vehicles traveling on each edge in the transportation network, which is then used to compute a travel time on each edge. The travel time on an edge is a function of the number of vehicles using that edge. These travel times are then given back to UrbanSim and ActivitySim to be used for congested accessibility and skims.

\subsection{Inputs}

The primary inputs used by the traffic assignment model comprise the network (including the associated origin-destination travel demand) and BPR coefficients associated with each edge. These inputs were described in detail in Sections \ref{sec:activitysim} and \ref{sec:network} and reviewed briefly here.

\subsubsection{The network}

The network infrastructure is based on a set of nodes and edges in which connected edges share a node. Vehicles are allowed to travel from one edge to another if the edges are connected via a node. The ordered set of these edges (in which the ordering denotes a shared node between two edges) are called a path. In order for a vehicle to move from its origin to its destination, it must take a particular path. The possible paths between origins and destinations are defined by the topology of the network (i.e., the number of edges, the number of nodes, and the connections between these nodes and edges). The network described in Section \ref{sec:network} is used as the foundation of the traffic assignment model.

\subsubsection{BPR coefficients}

The BPR coefficients computed previously are used to assign a pseudo travel time to each edge as a function of the edge's load (i.e. number of vehicles on the edge). This is necessary in order to determine which paths vehicles should be assigned to (since vehicles are more likely to take a path which requires less time).

\subsection{How it works}

The currently implemented version of the traffic assignment model determines a static user equilibrium using the Frank-Wolfe algorithm. Static traffic assignment does not consider time varying parameters of any kind so there is no concept of flow dynamics but there is a well-defined equilibrium. Static user equilibrium, often called Wardrop's first principle in the transportation literature, is a well-defined state in which all vehicles take the shortest path from their origin to their destination. The resulting traffic assignment is based on shortest paths that are calculated while considering the travel times resulting from a loaded network in which each user is attempting to minimize their travel time. 

Wardrop's first principle states that the actual travel time experienced by a user in the network is equal or less than the travel time that the same driver would experience on any other route \citep{wardrop1952road}. Static user equilibrium---which is equivalent to Nash equilibrium---is defined in Equation \ref{eq:static_user_equilibrium}:

\begin{equation}
    \label{eq:static_user_equilibrium}
    \begin{split}
    \min \sum_{e\in\ \textit{edges}}\int_{0}^{v_e} S_e(x) dx\\
    \text{subject to} \\ 
    v_e = \sum_i \sum_j \sum_r \alpha_{ij}^{er} x_{ij}^{r}\\
    \sum_r x_{ij}^{r} = T_{ij}\\
    v_e \geq 0 \\
    x_{ij}^{r} \geq 0
    \end{split}
\end{equation}

The minimization is over travel time on each edge as a function of traffic volume. $x_{ij}^{r}$ is the number of vehicles on path $r$ from node $i$ to node $j$. $\alpha_{ij}^{ar}$ is equal to 1 if edge $a$ is contained in path $r$ and 0 otherwise. To perform this static user equilibrium calculation we use the Frank-Wolfe algorithm \citep{frank1956algorithm}. The Frank-Wolfe algorithm is a first-order optimization algorithm that is used to solve convex problems such as the static user equilibrium problem defined above. The algorithm works as follows; consider a problem of the form:

\begin{align*}
    S \in \mathcal{R}^n \text{ a polyhedron and } f: \mathcal{R}^n \rightarrow \mathcal{R} C^1, \text{ solve:}
\end{align*}

\begin{align*}
    \text{min. } f(x) \text{ subject to } x \in S
\end{align*}

Frank-Wolfe algorithm:

\begin{enumerate}
    \item Initialize with $x_0 \in S$ and let $k= 0$
    \item Determine a search direction $d_k = y_k - x_k$ by solving the linear program: \\
    $y_k \in \text{argmin}_{y \in S} \{ \nabla f(x_k)^Ty \} $
    \item Determine a step length $\alpha_k \in [0.1]$ such that: \\
    $f(x_k + \alpha_k d_k) = f((1 - \alpha)x_k + \alpha y_k) \leq f(x_k)$
    \item Update $x_{k+1} = (1 - \alpha)x_k + \alpha y_k$ let $k = k+1$ and go to step 2
\end{enumerate}

\subsection{Outputs}

From the traffic assignment model, we obtain the number of vehicles on each edge in the network and the resulting travel time (calculated using the BPR coefficients) of each edge. Since these travel times are a result of a static traffic assignment, they are not interpretable as exact travel times and the units are effectively meaningless. Rather, they represent the congested travel time of each edge calculated using the BPR equations.

As described previously, these BPR function are empirical. These travel times are used to generate congested skims and accessibility, in which the congestion on each edge in the network is known in relative terms, and therefore areas of relatively high versus low congestion can be specified. UrbanSim and ActivitySim use these congested edge travel times to calculate accessibility and skims. For these calculations, the precise definition of travel time is not necessary; the comparison between edges is the important aspect.

\subsection{Calibration and validation}

In its current implementation, the traffic assignment model is static, meaning that back-propagation of congestion and other time dependent phenomena observed in real transportation networks cannot be well-modeled. However, the resulting travel times due to loading on each edge can be compared with real data to understand how well the static approximation matches with the observed travel times on a network. 

\subsection{Benchmarking}

Our current tests on traffic assignment at scale have been conducted on an SF Bay area network composed of $31,000$ edges, $66,000$ nodes, and a total network demand of $256,000$ vehicular trips. Given this network size and using the previously defined BPR functions and Frank-Wolfe algorithm, the static traffic assignment calculation is computed nearly instantaneously on a Windows machine with Intel Core i5 1.90Ghz CPUs.

\section{Traffic microsimulation model}
\label{sec:micro}
\subsection{Overview}

The traffic microsimulation component, in contrast to traffic assignment, simulates traffic using individual vehicles and a high detail representation of the network. This component simulates traffic interaction at the individual vehicle level including start time, speed, lane changing, and car following headways. It also simulates traffic controls such as stop signs and traffic lights. The traffic microsimulation component in this paper is an extension of work~\citep{garcia2014designing} to increase its computational scale using a large network.

A principal advantage of traffic microsimulation is that it allows creating and storing vehicle-specific information such as travel time, fuel consumption, trip routes, as well as aggregate information such as congested travel times, accessibility, network usage, and vehicle pollution.  Due to its high detail simulation of individual vehicles and the high detail representation of the transportation network, traffic microsimulation models are typically applied to small areas such as an intersection or a small highway corridor.  In that context, it may be surprising to consider using a traffic microsimulation approach to integrate with land use modeling and activity-based travel modeling for long-term (e.g. 30 years) simulations.  However, the implementation approach taken in this traffic microsimulation model uses a novel application on a graphics processing unit (GPU) enabling dramatic speedup from parallelization.  This component is still in early stages of development and testing as described here.

\subsection{Inputs}
The inputs used by the traffic microsimulation model comprise the network as described in section \ref{sec:network}
 (including an estimation of per road speed limits), the associated origin-destination travel demand, and the distribution of departure times to work in the morning, and to home in the afternoon.  The initial test application is only using commute trips to work for its demand, as an initial means of testing the computational approach.

\subsubsection{The network}
The network is based on a set of nodes (i.e., intersections) and directional edges (i.e., roads) where the nodes are connected through the edges. Vehicles start in one edge (defined by the origin) and traverse the network until it reaches the destination. The traversal is controlled by a per vehicle simulation (Subsection~\ref{sub:per_vehicle}) and the specific path (Subsection~\ref{sub:trip_planning}). The network described in Section~\ref{sec:network} is used as the foundation of the traffic microsimulation model. In addition to the nodes and edges, we need the average speed of each edge as well as the number of lanes in it (Section~\ref{sec:network}), these parameters directly influence the traffic behavior. Traffic controls at intersections are implemented in a very simplified way, initially, using procedural modeling.  

\subsubsection{OD and departure time}
Similar to Section~\ref{sec:ta}, we use the OD demand, described in Section~\ref{sec:activitysim} by ActivitySim, to provide the origin and destination of each vehicle. For each vehicle, we create a set of procedural variables (e.g., desired acceleration) following the suggested values by~\citep{treiber2000congested}.

\subsection{How it works}
The main components of our microsimulation component are trip planning (how to define each individual vehicle path), traffic atlas (a model to simulate traffic using the parallelization of the GPU), and per vehicle and traffic control simulation. The model is based on well-known traffic simulation literature and extended to work with the scale of this system.

\subsubsection{Trip Planning}\label{sub:trip_planning}
To create the individual trip planning of each vehicle we use as input the origin and destination from ActivitySim as well as the distribution of departure time. The microsimulation uses these to not just create the individual route but to simulate over time its individual behavior from the origin to the destination following each edge using a traffic model (Section~\ref{sub:per_vehicle}).

We approximate traffic equilibrium by using an iterative decaying approach. The idea is to initially create the paths using as estimated travel times for the given network, i.e., estimating the travel time ($t_t$) of each edge as $t_t= \sqrt{edge_{\# lines}} edge_{length}/edge_{speed}$. This causes most vehicles to use the highways and the shortest paths. However, since all vehicles choose those same edges when the traffic microsimulation is run, the estimated travel times differ due to the congestion. To approximate traffic equilibrium, we run several iterations of our traffic microsimulator. In the first iteration, we use the initialization formula above to compute the edge weight for routing.  After that first iteration, we use the microsimulation measurements to update those weights and repeat the microsimulation, so that the result is no longer a deterministic function of the initialization equation, and reflects a behavioral adaptation by vehicles adapting to congestion. After four iterations in our examples, the system converged and the estimated travel times match the simulated ones. Note that in order to dampen fluctuations that would occur between iterations, we do not recompute the path for each vehicle in each iteration but select randomly an increasing subset of vehicles that do not alter it. In our implementation, we recompute all routes the first and second iteration, and then we decay it to half and one-fourth of the vehicles.

\subsubsection{Per-vehicle and traffic control simulation}\label{sub:per_vehicle}
Given the set of per-vehicle paths, we simulate the traffic by updating each individual vehicle: lane changing (mandatory or discretionary) and vehicle following, and update the travel time, position, and velocity. Each update step is simulated over many small time steps (e.g., $\Delta t \in [0.1,1.0] s$). In our implementation, we use $\Delta=0.5 s$ that yields accurate simulation for speeds up to $100mph$. Note that despite using such a small time step, the simulation models can be simpler and parallelizable yielding higher throughput than other approaches.

\begin{itemize}
  \item \textit{Car-Following Model:} The current acceleration of a vehicle depends on the proximity to the \textit{following car}, i.e., if there is not a following car the vehicle will try to reach the free flow speed (maximum speed of the current edge), otherwise, it will adapt the acceleration to keep a safe distance to the following car. We use the Intelligent Driver Model~\citep{treiber2000congested} as:
  
\begin{equation}\label{eq:car-follow}
\dot{v}=a \left( 1- ( v/v_o )^{4}- ((s_{0}+Tv+v\Delta v/2\sqrt{ab})/s )^{2} \right)
\end{equation}

where $a$ is the acceleration ability of the vehicle, $v$ is the current speed of the car, $v_0$ is the edge speed limit, $s$ is the distance gap to the following car, $s_{0}$ is the minimum following distance, $T$ is the desired time headway, and $b$ is a comfortable braking deceleration of the vehicle.

  \item \textit{Lane-Changing Model:} Besides the car position, velocity and acceleration update, each car might perform a lane-change~\citep{choudhury2007modeling}. This follows two main behaviors: i) mandatory behavior (necessary change to reach an exit or right/left turn) or ii) discretionary (lane change to increase speed and bypass a vehicle). When a vehicle enters a road, it starts with discretionary behavior and it changes of state (if it was not in the correct lane) to mandatory exponentially with the reminding distance to the exit ($e^{-(x_i-x_0)^2}$ where $x$ is the current distance to the exit and $x_0$ is the last exit position).
  \item \textit{Gap-Acceptance Model.} Once the vehicle has decided to change lanes, the maneuver is performed if the lead and lag gaps are acceptable.
\end{itemize}

Traffic controls at intersections are implemented in a very simplified way, initially, using procedural modeling. Our system implements stop and traffic light signals. For stop controls our system allows just one vehicle at a time into the intersection and waits until the car has reached the next edge. For traffic lights, given an intersection with $n$ inbound edges and $m$ outbound, our system creates the geometric combination $nm$ of necessary traffic phasing to allow in a round robing logic to all cars pass the intersection. We plan to extend this initial simplified intersection logic using real-world data to infer the type of intersection and phasing.

\subsubsection{Traffic atlas}
To keep all the variables described above and efficiently compute distances between vehicles (e.g., $s$ and lead and lag gaps), we use the concept of a  \textit{traffic atlas}. Instead of keeping a list of vehicles per lane that would not be parallelizable, we represent a road as a set of contiguous bytes in memory, in order to find the leading car of a given vehicle, we check the following bytes in memory. This approach allows us to use the many cores on a graphics card to parallelize its processing.

\subsubsection{Vehicle energy consumption and pollution estimates}

Our simulator, since it has the position, speed, acceleration of each vehicle at every given time, can be used to calculate vehicle gas consumption and pollution estimations. In the current implementation, we compute gas consumption as well as CO emissions but this can be extended to compute any kind of pollutant such as nitrous oxide ($N_2O$), methane ($CH_4$), or hydrofluorocarbons ($HFC$) emissions~\citep{abou2013using, ahn2002estimating}. 

To report total or per-vehicle CO emission, our system uses the following approximation for the measured emissions~\citep{stein2003link} as 
\begin{equation}\label{eq:pollution}
{P = -0.064 + 0.0056v_m + 0.00026(v_m-50)^2}
\end{equation}
where $P$ is the emission rate (in grams of CO per second) and $v_m$ is the speed of the vehicle in miles per hour.

To compute the gas consumption, we use the approximation proposed by \cite{akcelik1989efficiency}. Every second we update the current gas consumption using the formula:
\begin{equation}\label{eq:consumtion}
f=0.666+0.072\left(0.269v+ 0.0171v^2+0.000672v^3 +1.680av+P_{ea} \right)
\end{equation}
where $f$ is the instant fuel consumption rate measured in $mL/s$ and $P_{ea}$ is the extra engine drag and computed as $P_{ea}=0.79296a^2v$ for $a>0$.

\subsubsection{Simulation scale}
Most microsimulation systems and the one we use as foundation for the current traffic microsimulator~\citep{garcia2014designing} were designed to handle few hundreds of km of roads and thousands of edges, this project tries to address is a network of over 66000 edges, 131000 road lanes, and 31Mm of roads. Moreover, the goal is to simulate the population of the 9 Bay Area counties that contains over seven million people. In short, this full scale of the San Fransisco Bay Area 9-county region is so large that former approaches are incapable of simulating it or would be too slow to be useful in practice.

The main changes to our initial microsimulator are:
\begin{itemize}
    \item Given the number of vertices ($V$), edges ($E$), and agents/vehicles ($A$), most approaches use the well-known algorithm Dijkstra~\citep{dijkstra1959note} that has a complexity of $\mathcal{O}(V log V)$. Since $A$ is a much bigger than $V$ this produces a prohibitive complexity. Therefore, we replace it with the also well-known algorithm \textit{All Pairs Shortest Path} \citep{johnson1977efficient} that has a complexity of $\mathcal{O}(VE log V)$. Since $E<<A$, using this approach we speed up the computation of paths in each iteration by two orders of magnitude.
    \item Given the number of road lanes and the maximum length of all roads, with our previous approach, we would have required the order of the order of dozens of GB of RAM memory to represent the traffic atlas. In this new version, we change the memory layout such that each road is discretized each 1000m making the memory to be reduced to the order of 270MB.
    \item We improved the generation of agents (i.e., vehicles) in the system. We included a second type of vehicle to represent trucks, to reflect their differences in size, acceleration, and maximum speed, and how these alter traffic patterns.
\end{itemize}

\subsection{Outputs}

\begin{figure}[bt]
    \center
    \includegraphics[width=\textwidth]{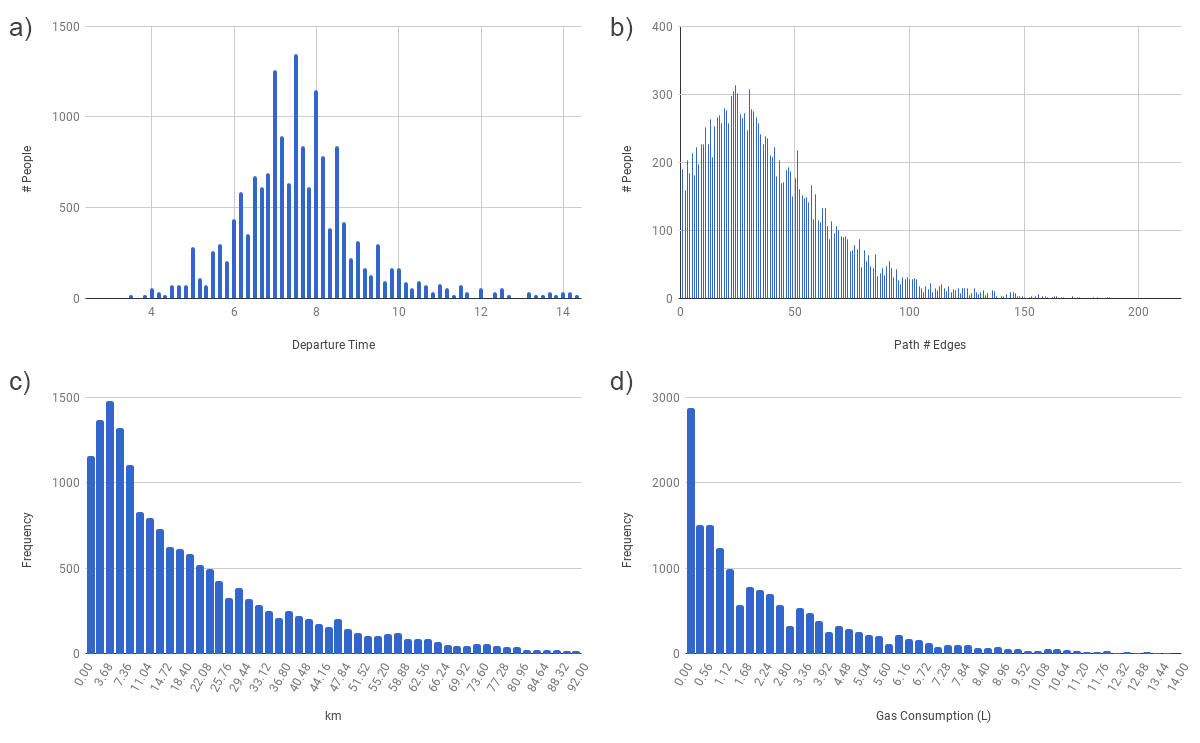}
    \caption{Microsimulation. a) Frequency of departure times; b) Simulation departure times; c) Histogram of the road length in the network; d) Histogram of number of people respect the number of edges in their path; e) Histogram of distance travel per vehicle; f) Histogram of the gas consumption per vehicle.}
    \label{fig:traffic-figure}
\end{figure}

\begin{figure}[tbhp]
    \center
    \includegraphics[width=0.8\textwidth]{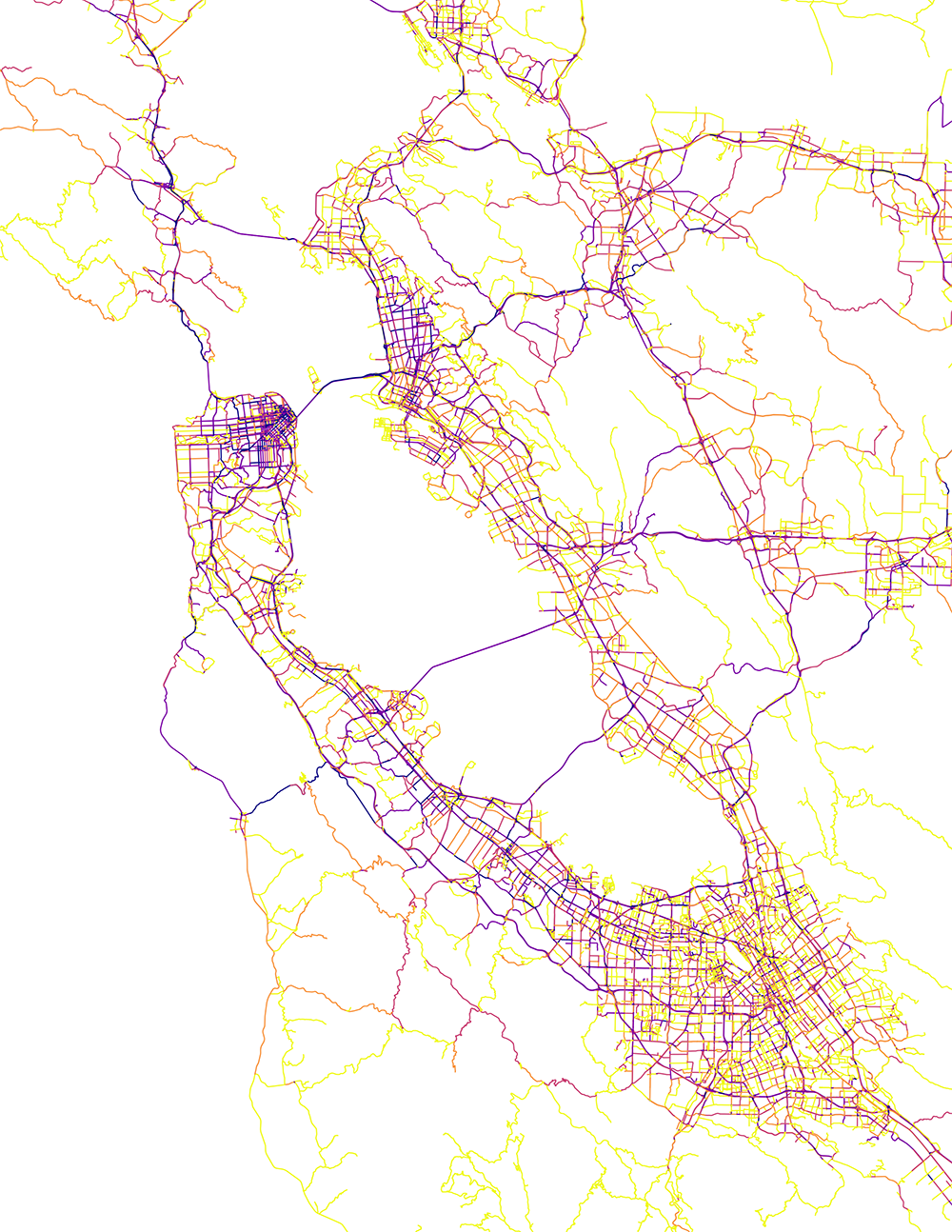}
    \caption{Relative average utilization of edges in the center of the study area, from low (yellow) to high (purple).}
    \label{fig:edge_avg_util}
\end{figure}

From the traffic microsimulation model, we sample every minute the utilization and per edge speed and average in groups of 6 minutes to generate 10 samples per hour. As initial microsimulation result, we run the network with the OD created by ActivitySim and run four iterations of the simulation to converge. Given we simulate individual vehicles, we can extract many measurements:
\begin{itemize}
    \item \textit{Departure times:} Figure~\ref{fig:traffic-figure}.a shows the histogram of departure times used for the traffic simulator. This was computed from the frequency of departure time histogram, first selecting a bucket by randomly matching the frequencies, and then by adding a homogeneous random variable scaled to the bucket time span.
    \item \textit{Number of edges distribution}. Using our iterative trip planning, we can converge to an approximation of equilibrium and analyze the number of edges that each person has to drive. As depicted by Figure~\ref{fig:traffic-figure}.b, most vehicles traverse less than 45 edges.
    \item \textit{Histogram of driven distance and gas consumption:} Figure~\ref{fig:traffic-figure}.c depicts the frequency of the distance driven per vehicle, most vehicles do short trips. Figure~\ref{fig:traffic-figure}.d shows the histogram of the gas consumption in liters, there is a correlation with the distance but time in traffic jams and traffic intersections increases the gas consumption.
    \item \textit{Utilization and Speed:} Figure~\ref{fig:edge_avg_util} shows the utilization of the road network. We defined utilization as the number of vehicles at the sampling time divided by the maximum number of vehicles that can fit on the edge. This capacity is computed as the length of the edge times the number of lines divided by the average length of a vehicle plus the lead gap.
    
\end{itemize}

\subsection{Calibration and validation}

Calibration of any microsimulation component is one of the most complicated tasks. From the simulation perspective, our traffic microsimulator was validated by comparing it against \textit{SUMO}. For the same scenario, both produced in average within 6\% occupancy values.
We will use the traffic assignment (Section~\ref{sec:ta}) results as comparative as well as real data respect to travel times and street utilization.

\section{Discussion and Conclusion}
\label{sec:conclusion}

This paper presented the preliminary architecture of an integrated modeling pipeline that joins together long-term land use, short-term travel demand, and modular traffic models including static user equilibrium assignment and vehicle microsimulation.


This ongoing effort will focus on adding workplace choice and vehicle ownership into the UrbanSim models and tightly integrating this with network models. With the behaviorally integrated models and a sufficiently detailed representation of the transportation network and geography, ideally using local streets and parcels and buildings, we will be able to explore the long-term feedback effects of transportation infrastructure changes on urban development patterns. We will be able to include the feedback effects of these urban development dynamics on travel demand and on travel flows and speeds, and consequently on transportation-related energy consumption.




By creating the combined models with sufficient performance to simulate the feedback of land use, travel, and congestion annually rather than every 5 to 10 years, a variety of alternative scenarios can be developed and explored that couple transportation and land use policies. Examples include transit-oriented development with transportation networks emphasizing transit, or more efficient highway projects coupled with supporting land use policies, with evaluation of the induced demand effects of both. Policies that affect parking construction and management, or which begin to explore the potential impacts of greater adoption of ride hailing services and eventually autonomous vehicles, could also be incorporated as data availability permits.

Finally, our research agenda includes substantial benchmarking and validation of this modeling pipeline. The benchmarking will measure run-time for various configurations of the models, with a research objective of improving run-time performance particularly for the travel demand model. We will also be exploring trade-offs between model granularity and representation versus computational performance. The validation step will compare our model's outputs with real-world data to verify plausible results.



\clearpage
\addcontentsline{toc}{section}{Notes}
\theendnotes

\clearpage
\bibliographystyle{apalike}
\bibliography{references}

\end{document}